\documentclass{aa}
\usepackage{graphicx}
\usepackage{txfonts}

\begin{document}

  \title{Optical identification of the companion to PSR~J1911$-$5958A,
    the pulsar binary in the outskirts of NGC\,6752}

  \author{
    C. G. Bassa\inst{1}
    \and F. Verbunt\inst{1}
    \and M. H. van Kerkwijk\inst{2}
    \and L. Homer\inst{3}
  }
  
  \institute{Astronomical Institute, Utrecht University, PO Box 80\,000, 3508 TA
    Utrecht, The Netherlands \and
    Dept. of Astronomy and Astrophysics, Univ. of Toronto, 60 St.
    George Street, Toronto, ON M5S 3H8, Canada \and
    Dept. of Astronomy, Univ. of Washington, Box 351580, Seattle, WA
    98195-1580, USA
  }

  \offprints{C.G. Bassa,\\ \email{c.g.bassa@astro.uu.nl}}

  \date{Received / Accepted}

  \abstract{We report on the identification of the optical counterpart
    of the binary millisecond pulsar PSR~J1911$-$5958A, located in the
    outskirts of the globular cluster NGC\,6752. At the position of
    the pulsar we find an object with $V=22.08$, $B-V=0.38$,
    $U-B=-0.49$. The object is blue with respect to the cluster main
    sequence by 0.8 magnitudes in $B-V$. We argue that the object is
    the white dwarf companion of the pulsar. Comparison with white
    dwarf cooling models shows that this magnitude and colors are
    consistent with a low-mass white dwarf at the distance of
    NGC\,6752. If associated with NGC\,6752, the white dwarf is
    relatively young, $\la\!2\,$Gyr, which sets constraints on the
    formation of the binary and its ejection from the core of the
    globular cluster.

  }

  \titlerunning{Optical identification of the companion to PSR~J1911$-$5958A}

  \maketitle

  \section{Introduction}
  Recently, 5 millisecond pulsars have been discovered (D'Amico et
  al. \cite{damico_c}) towards the nearby galactic globular cluster
  NGC\,6752. Three of them are located inside the $6\farcs7$ core
  radius (Lugger et al. \cite{lugger}) while the other two are outside
  the $1\farcm92$ half-mass radius (Djorgovski \cite{djorgovski}), at
  $2\farcm7$ and $6\farcm4$, respectively. The latter of these is a
  binary millisecond pulsar, PSR~J1911$-$5958A (hereafter PSR~A), and
  has a low-mass ($\ga\!0.19\,\mathrm{M}_\odot$, assuming a
  $1.4\,\mathrm{M}_\odot$ neutron star) companion in a 20~hour, highly
  circular ($e<10^{-5}$) orbit (D'Amico et al. \cite{damico_c}). The
  pulsar period and period derivative suggest that it is a
  ``canonical'' recycled millisecond pulsar (see Phinney \& Kulkarni
  \cite{phinney} for a review), and hence that the companion is likely
  a white dwarf.

  The large separation of PSR~A from the cluster center is
  puzzling. Colpi et al. (\cite{colpi}) have recently investigated
  possible scenarios, and found that both for the case of a primordial
  binary and for an exchange or scattering event with other cluster
  stars, it is very difficult to explain both the pulsar's current
  position and its close circular binary orbit. Instead, they suggest
  that the binary may have been scattered to its current position by a
  binary composed of two 3--100\,$\mathrm{M}_\odot$ black holes.

  One might learn more about the system's origin (and verify cluster
  membership) if one can confirm that the companion is a white dwarf
  and measure its mass and age. Therefore, we searched for the optical
  counterpart in archival data. We report here on the results.

  \section{Observations and analysis}\label{sec:observations}
  We searched the ESO and \emph{Hubble Space Telescope} archives for
  images coincident with the pulsar position
  ($\alpha_\mathrm{2000}=19^\mathrm{h}11^\mathrm{m}42\fs7562$,
  $\delta_\mathrm{2000}=-59\degr58\arcmin26\farcs900$; D'Amico et
  al. \cite{damico_c}). We found two images, taken with the Wide Field
  Imager (WFI) at the ESO 2.2~m telescope on La Silla. These
  observations, 4~minute $B$ and $V$-band exposures, were taken during
  the night of May 13/14, 1999. The seeing was poor, $\sim\!1\farcs5$
  in $V$. However, both images showed a faint object at the pulsar
  position (see below). This object was also present in two \emph{HST}
  observations with the Wide Field Planetary Camera 2 (WFPC2; Holtzman
  et al. \cite{holtzman_a}). Both observations, \texttt{U5FI07} and
  \texttt{U5FI03} (GO-8256), were imaged in the same filters and had
  similar exposure times, 42~s in F555W (hereafter $V_{555}$), 220~s
  in F439W ($B_{439}$), 660~s in F336W ($U_{336}$) and 1\,800~s and
  1\,693.5~s in F255W ($\mathrm{nUV}_{255}$) for the first and second
  field, respectively. The position of the pulsar coincides with the
  WF3 chip for the \texttt{U5FI07} dataset, while it is on the WF4
  chip on the other dataset.

  \subsection{Astrometry}
  The WFI detector has an array of 8 CCDs (2 rows of 4), each CCD
  having a field of view of $8\arcmin\times16\arcmin$, a total of
  $33\arcmin\times34\arcmin$. The position of the pulsar was
  coincident with chip 6 of the $V$-band image. We found that there
  was some distortion over the whole chip. To minimize its effect we
  only used the upper half of this chip for the astrometric
  calibration. Stars on this $8\arcmin\times8\arcmin$ sub-image were
  compared against entries in the USNO CCD Astrograph Catalog (UCAC;
  Zacharias et al. \cite{zacharias}). In total 68 UCAC stars coincided
  with this image and their centroids were measured. Of these 43 were
  not saturated and appeared stellar and unblended. One outlier,
  having a total residual of $0\farcs32$ was rejected. The remaining
  stars were used to calculate an astrometric solution, fitting for
  zero-point position, scale and position angle. This solution has
  root-mean-square (\emph{rms}) residuals of $0\farcs06$ in both right
  ascension and declination. The statistical uncertainty in the
  astrometry is thus $0\farcs084$ in each coordinate.

  We used the astrometrically calibrated WFI image to obtain
  astrometric solutions for the two \emph{HST}/WFPC2 datasets. First,
  the WFPC2 pixel positions were corrected for geometric distortion
  and placed on a master-frame, using the prescription of Anderson \&
  King (\cite{anderson}). We matched stars on the WFI image and fitted
  for zero-point position, scale and position angle against the WFPC2
  master frame positions. Outliers having residuals larger than three
  times the \emph{rms} residual of the fit are removed and a new
  solution is computed. This process is iterated until convergence.
  On average the converged astrometric solution used some 200--300
  stars with \emph{rms} residuals of the order of $0\farcs03$ in both
  right ascension and declination. The final uncertainty in the tie to
  the UCAC system is dominated by the step from the UCAC to the WFI
  sub-image, and is $\sim\!0\farcs10$ in each coordinate.

  \begin{figure}
    \resizebox{\hsize}{!}{\includegraphics{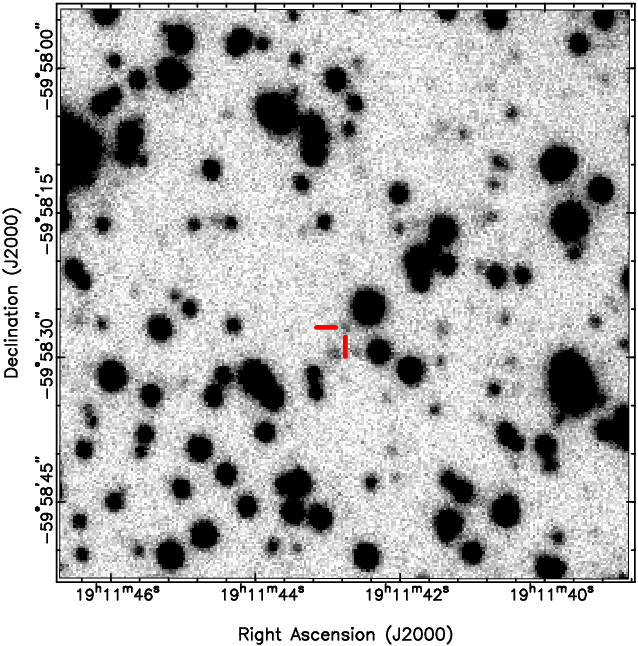}}
    \resizebox{\hsize}{!}{\includegraphics{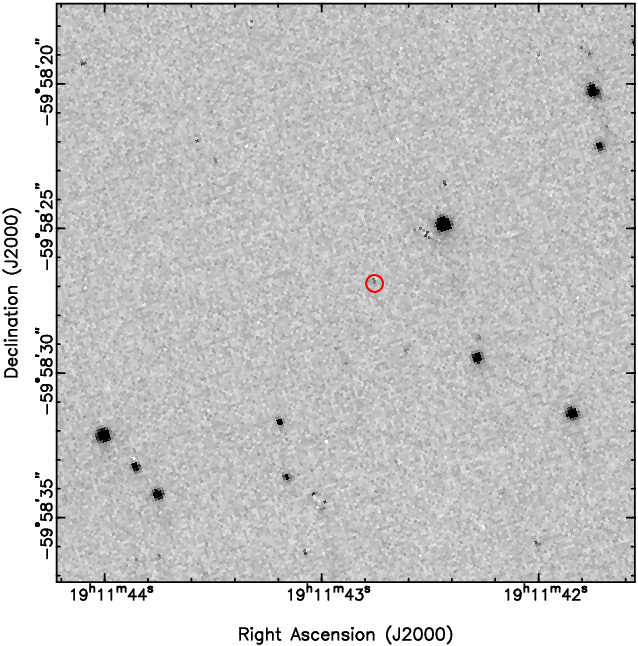}}
    \caption{Finding charts for PSR~A. The upper image is a
      $1\arcmin\times1\arcmin$ subsection of a $V$-band image obtained
      on May 14,1999 with the Wide Field Imager on the ESO 2.2~m
      telescope at La Silla. The $2\arcsec$ tick marks indicate the
      position of PSR~A. The lower image is a
      $20\arcsec\times20\arcsec$ sub-image of the \emph{HST}/WFPC2
      $V_{555}$ (F555W) observation. The $0\farcs235$ (95\% confidence)
      error circle is shown at the position of the pulsar.}
    \label{fig:image}
  \end{figure}

  \subsection{Photometry}
  We started with the pipe-line calibrated \emph{HST}/WFPC2 images,
  and used the HSTphot 1.1 (Dolphin \cite{dolphin_a}) package for
  further reduction and photometry of the images. We followed the
  recommended procedures to mask bad pixels, defects, cosmic ray hits
  and hot pixels. Next we used the main task \emph{hstphot} to find
  stars, measure positions, and determine calibrated photometry. The
  latter uses the aperture corrections, charge-transfer efficiency
  corrections and zero-points of Dolphin (\cite{dolphin_b}).

  \begin{table*}
    \centering
    \caption[]{Positional and photometric data.}
    \label{tab:data}
    \begin{tabular}{llllcccc}
      \hline
      Dataset & Date (UT) & R.A. (J2000) & Decl. (J2000) &
      $\mathrm{nUV}_{255}$ & $U_{336}$ & $B_{439}$ &
      $V_{555}$ \\
      \hline
      \texttt{U5FI07} & March 2, 2000 @ 10:29 &
      $19^\mathrm{h}11^\mathrm{m}42\fs756$ &
      $-59\degr58\arcmin26\farcs87$ & $20.74\pm0.23$ &
      $21.49\pm0.15$ & $22.40\pm0.21$ & $22.10\pm0.17$ \\
      \texttt{U5FI03} & March 3, 2000 @ 13:54 &
      $19^\mathrm{h}11^\mathrm{m}42\fs757$ &
      $-59\degr58\arcmin26\farcs83$ & $21.44\pm0.24$ &
      $21.78\pm0.14$ & $22.49\pm0.17$ & $22.09\pm0.14$ \\
      \hline
    \end{tabular}
  \end{table*}

  \subsection{The counterpart to PSR~J1911$-$5958A}
  The UCAC catalog is on the ICRS at the 20 mas level (Assafin et
  al. \cite{assafin}). Including this uncertainty in the uncertainty
  of the astrometric tie, we obtain 95\% confidence radii for the WFI
  and \emph{HST}/WFPC2 frames of $0\farcs211$ and $0\farcs235$,
  respectively. Within these radii there is a single object, see
  Fig.~\ref{fig:image}. The \emph{HST}/WFPC2 positions and magnitudes
  are tabulated in Table~\ref{tab:data}.
  
  The photometric measurements of the object in the two datasets are
  consistent for all filters except $\mathrm{nUV}_{255}$. The object
  is 0.7~magnitudes brighter in the first dataset than in the
  second. However, we compared the magnitudes of 240 stars that
  overlapped between the two datasets and found that, on average,
  stars in the first dataset were brighter by 0.34~magnitudes in
  $\mathrm{nUV}_{255}$, which removes the observed discrepancy. We
  have not, however, found an explanation for this large offset
  between the two datasets.

\begin{figure}
    \resizebox{\hsize}{!}{\includegraphics{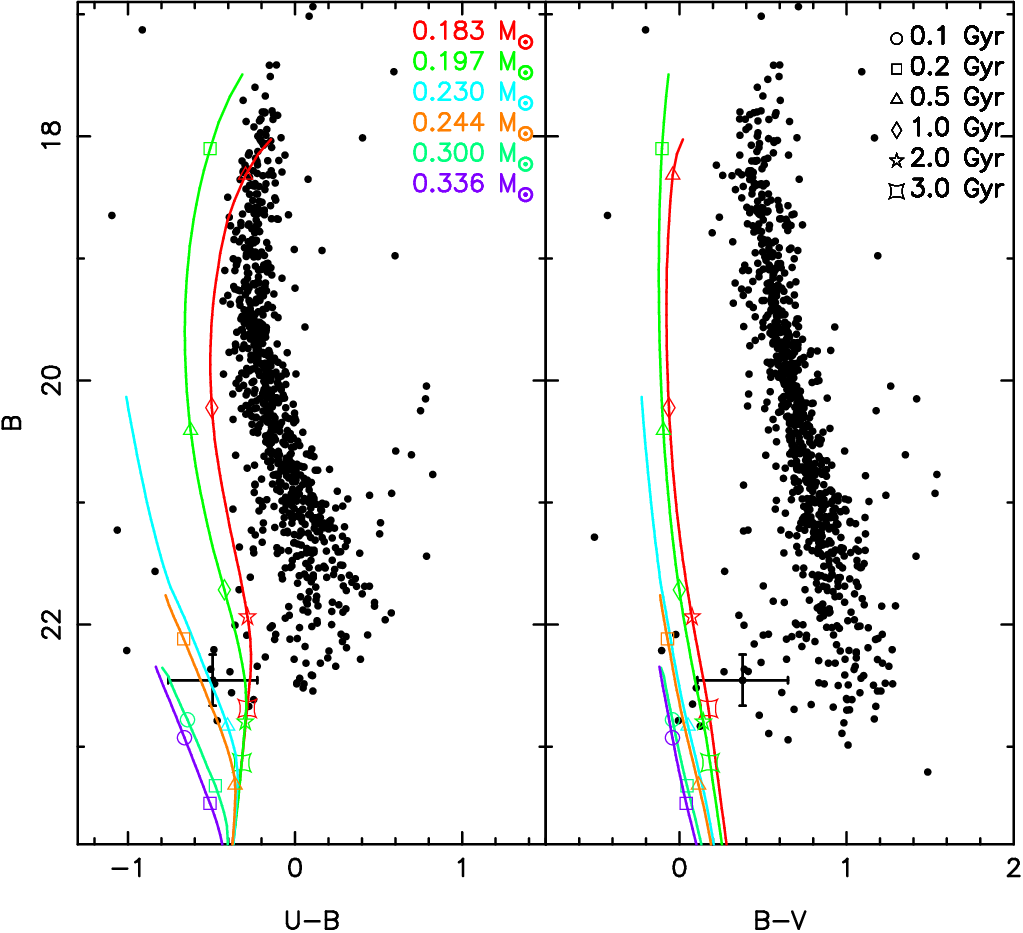}}
    \caption{Color-magnitude diagrams of the two \emph{HST}/WFPC2
      fields. The magnitudes are transformed from the \emph{HST}/WFPC2
      flight system to Johnson-Cousins $UBV$ using the prescription by
      Holtzman et al. (\cite{holtzman_b}). The average colors and
      magnitudes of the counterpart to PSR~A are indicated with error
      bars. The errors include the uncertainty in the distance modulus
      and reddenning. Also shown are Helium--core white dwarf cooling
      tracks (solid lines) by Serenelli et al. (\cite{serenelli}). The
      models have masses of 0.183, 0.197, 0.230, 0.244, 0.300 and
      $0.336~\mathrm{M}_\odot$, decreasing in mass to the red, for a metallicity of
      $Z=0.001$. The age of the white dwarfs along the cooling track is
      indicated.}
    \label{fig:cmds}
  \end{figure}

  The magnitudes in the \emph{HST} flight system filters ($U_{336}$,
  $B_{439}$ and $V_{555}$) were transformed to the Johnson-Cousins
  $UBV$ (Vega) system using the transformations by Holtzman et
  al. (\cite{holtzman_b}). The resulting average magnitudes for the
  object in this system are $U=21.96\pm0.10$, $B=22.46\pm0.14$ and
  $V=22.08\pm0.11$. The object is blue by about 0.8~magnitudes with
  respect to the cluster main sequence in $B-V$ at the same $B$-band
  magnitude, Fig.~\ref{fig:cmds}. This is also confirmed by an
  instrumental color-magnitude diagram constructed from the WFI $B$
  and $V$-band images.

  Given the density of objects on the \emph{HST}/WFPC2 chips, there is
  a $\sim\!1$\% probability of a chance coincidence in the 95\%
  confidence \emph{HST} error circle. However, there are only a few
  stars that are blue with respect to the cluster main sequence, which
  greatly reduces the probability of a chance coincidence. The blue
  color, together with the positional coincidence, gives us confidence
  that we have detected the companion of PSR~A.

  \section{Ramifications}\label{sec:ramifications}
  The minimum mass for the companion is constrained by the pulsar mass
  function. For a pulsar mass of $1.35~\mathrm{M}_\odot$ (Thorsett \&
  Chakrabarty \cite{thorsett}) this minimum mass is
  $0.185~\mathrm{M}_\odot$. The lower limit increases for heavier
  pulsars, roughly by $0.004~\mathrm{M}_\odot$ for every
  $0.05~\mathrm{M}_\odot$ step in the pulsar mass. Assuming a random
  probability distribution for the inclination of the binary, we find
  that there is 90\% probability that the companion mass is less than
  $0.5~\mathrm{M}_\odot$. Given this range of masses and the fact that
  PSR~A is a recycled millisecond pulsar, it is likely that the
  companion is a Helium-core white dwarf.

  To verify whether our observations are compatible with a Helium-core
  white dwarf at the distance of the cluster, we compare our
  magnitudes with the predictions from the white dwarf cooling tracks
  of Serenelli et al.\ (\cite{serenelli}). We use their tracks for
  $Z=0.001$, as this metallicity provides the best match to the
  metallicity of NGC\,6752 ($\mbox{[Fe/H]}=-1.43\pm0.04$; Gratton et
  al.\ \cite{gratton}). We also assume a $V$-band distance modulus
  $(m-M)_V=13.24\pm0.08$ and reddening $E_{B-V}=0.040$, as recently
  determined by Gratton et al.\ (\cite{gratton}), and use the relative
  extinction coefficients listed by Schlegel et al.\
  (\cite{schlegel}).

  Figure~\ref{fig:cmds} shows the cooling tracks for Helium-core white
  dwarfs with $Z=0.001$ for masses in the range of 0.183 to
  $0.336~\mathrm{M}_\odot$. It appears that the magnitude and colors
  of the companion to PSR~A are compatible with the two lowest-mass
  tracks, 0.183 and $0.197~\mathrm{M}_\odot$. Note that the
  $0.183~\mathrm{M}_\odot$ model is below the minimum mass inferred
  from the pulsar mass function.

  We have fitted the observed absolute $UBV$ magnitudes against the
  predictions from the $Z=0.001$ Helium-core white dwarf models. A
  $\chi^2$ statistic was computed for each entry in the model from the
  difference between the observed and modelled absolute
  magnitudes. Table~\ref{tab:fitting} shows, for each model with a
  given mass, the properties of the white dwarf at the $\chi^2$
  minimum.  Both Fig.~\ref{fig:cmds} as Table~\ref{tab:fitting} show
  that the lowest mass models, 0.197 to 0.244~$\mathrm{M}_\odot$, are
  preferred. At these masses the white dwarf is rather hot, with
  $T_\mathrm{eff}\approx11\,000-16\,000$~K.

  Given these high temperatures, the counterpart is relatively young,
  $\la\!2$~Gyr (see both Fig.~\ref{fig:cmds} and
  Table~\ref{tab:fitting}). The precise value strongly depends on the
  mass, since white dwarfs with lower masses have relatively thick
  hydrogen envelopes, where residual hydrogen shell burning keeps the
  white dwarf hot. From the values listed in Table~\ref{tab:fitting},
  one sees a jump in the cooling age between the 0.197 and
  $0.230~\mathrm{M}_\odot$ models. This reflects a dichotomy in the
  thickness of the hydrogen layer, where, above a certain critical
  mass, the thickness has been reduced by shell flashes early in the
  evolution (Althaus et al. \cite{althaus}). As the shell burning is
  through the CNO cycle, the critical mass depends on the metallicity:
  $\sim\!0.18~\mathrm{M}_\odot$ for solar metallicity,
  $\sim\!0.22~\mathrm{M}_\odot$ for $Z=0.001$ and
  $\sim\!0.26~\mathrm{M}_\odot$ for $Z=0.0002$ (Althaus et
  al. \cite{althaus}; Serenelli et al. \cite{serenelli}).

  \begin{table}
    \centering
    \caption[]{Fitting results for Helium--core white dwarf models
      with a metallicity of $Z=0.001$. The observed absolute $UBV$
      magnitudes were fitted against the modelled values. The
      dereddened, observed values $U-B$, $B-V$ and $M_\mathrm{V}$ are
      given in the second row.}
    \label{tab:fitting}
    \begin{tabular}{lcccccc}
      \hline
      Mass & $T_\mathrm{eff}$ & $\tau_\mathrm{c}$ & $(U-B)_0$ &
      $(B-V)_0$ & $M_\mathrm{V}$ & $\chi^2_\nu$ \\
      ($\mathrm{M}_\odot$) & (kK) & (Gyr) & -0.54 & 0.34 & 8.84
      & \\
      \hline
      0.183 & 10.6 & 2.30 & -0.31 &  0.08 &  8.92 &  0.9\\
      0.197 & 11.7 & 1.36 & -0.37 &  0.03 &  8.98 &  1.1\\
      0.230 & 14.7 & 0.37 & -0.59 & -0.06 &  9.14 &  3.4\\
      0.244 & 15.6 & 0.27 & -0.64 & -0.08 &  9.18 &  4.2\\
      0.300 & 19.1 & 0.07 & -0.80 & -0.13 &  9.29 &  8.1\\
      0.336 & 20.3 & 0.05 & -0.85 & -0.14 &  9.33 &  9.5\\
      0.380 & 21.6 & 0.05 & -0.89 & -0.16 &  9.34 & 10.6\\
      0.390 & 22.1 & 0.04 & -0.91 & -0.16 &  9.34 & 11.1\\
      0.422 & 23.8 & 0.03 & -0.95 & -0.18 &  9.39 & 12.9\\
      0.449 & 24.7 & 0.06 & -0.98 & -0.19 &  9.40 & 13.7\\
      \hline
    \end{tabular}
  \end{table}

  We should note that the dichotomy in cooling age was not found by
  Driebe et al. (\cite{driebe_a}), who found the flashes hardly
  affected the thickness of the hydrogen layer. As a result, more
  massive white dwarfs cool slower and are older. However, for the
  mass range of interest here, their results are similar. Their
  $0.195~\mathrm{M}_\odot$ model has a cooling age of 1.2~Gyr,
  comparable to the age of the Serenelli $0.197~\mathrm{M}_\odot$
  model. For higher masses the cooling age decreases. The
  $0.300~\mathrm{M}_\odot$ model by Driebe et al. (\cite{driebe_a})
  has a cooling age of 0.2~Gyr, down to 25~Myr for the heaviest model,
  with a mass of $0.414~\mathrm{M}_\odot$.
  
  Finally, we note the similarity between the position in the
  color-magnitude diagram of the companion of PSR~A and the companion
  of \object{PSR~J0024$-$7204W} (Edmonds et al. \cite{edmonds_b}) in
  \object{47\,Tucanae}. On the basis of optical variability with the
  orbital period (3.2~h) and eclipses of the pulsar radio emission,
  the companion is argued to be a heated main sequence star. For
  PSR~A, however, we can exclude this possibility: the spin-down
  luminosity is too low to lead to heating to the observed
  temperature.

  \section{Discussion and conclusions}
  We have detected the optical companion to the binary millisecond
  pulsar PSR~J1911$-$5958A, which is located $6\farcm4$ from the
  center of NGC\,6752.  The companion is blue with respect to the
  cluster main sequence by 0.8 magnitudes in $B-V$ and comparison of
  its colors and magnitude with white dwarf models shows that it is
  consistent with a Helium-core white dwarf at the distance of
  NGC\,6752.

  Irrespective of the cooling models we use, we find that the white
  dwarf is at most 2~Gyr old. This age is similar to the
  $\ga\!0.7$~Gyr the binary can be expected to stay in the outskirts
  if it is currently on a highly eccentric orbit in the cluster (Colpi
  et al. \cite{colpi}), which suggests that the white dwarf formed
  during, or shortly after, an encounter that also ejected the binary
  from the core, in an exchange interaction involving a binary with
  another star or binary. In the scenarios in which the binary was
  formed in the periphery, or scattered by a binary black hole, the
  coincidence of the two time scales has to be due to chance.

  One would expect the characteristic age of the pulsar to be similar
  to the cooling age of the white dwarf, as the pulsar starts to spin
  down after the cessation of mass transfer, while the white dwarf
  starts to cool. This seems not to be the case for PSR~A, for the
  characteristic age of the pulsar is $P/(2\dot{P})\sim\!17$~Gyr
  (D'Amico et al. \cite{damico_c}). It may be instead that the
  assumption underlying the characteristic age is wrong, and that the
  period at which the pulsar started spinning is similar to the
  current one.

  Compared to other white dwarf companions to milliseconds pulsars,
  the companion to PSR~J1911$-$5958A is bright, $V\simeq22$, which
  opens the possibility to determine detailed physical parameters
  (e.g., Van Kerkwijk et al. \cite{vankerkwijk}, Callanan et
  al. \cite{callanan}). For instance, one could measure the white
  dwarf temperature and surface gravity through spectroscopy. By
  comparison with models, this would lead to a mass and radius of the
  white dwarf, as well as a much more precise cooling age. The radius
  would allow one to confirm the association of the binary with
  NGC\,6752. If the association is confirmed, the more accurate
  distance to NGC\,6752 provides additional constraints on the radius
  and thus the mass of the white dwarf. Combining the white dwarf mass
  with a radial-velocity orbit of the white dwarf (and thus a mass
  ratio), would give the mass of the pulsar.

  \begin{acknowledgements}
    This research is based on observations made with the European
    Southern Observatory telescopes obtained from the ESO/ST-ECF
    Science Archive Facilities and observations made with the NASA/ESA
    \emph{Hubble Space Telescope}, obtained from the data archive at
    the Space Telescope Institute. STScI is operated by the
    association of Universities for Research in Astronomy, Inc. under
    NASA contract NAS 5-2655. CGB is supported by the Netherlands
    Organization for Scientific Research.
  \end{acknowledgements}

\end{document}